\documentclass[12pt]{iopart}

\usepackage{iopams}
\usepackage{graphicx}
\usepackage{amsfonts}

\begin{document}

\title{Quantum counterpart of energy equipartition theorem for fermionic systems}

\author{Jasleen Kaur, Aritra Ghosh \& Malay Bandyopadhyay}

\address{School of Basic Sciences,\\ Indian Institute of Technology Bhubaneswar, Argul, Jatni, Khurda, Odisha 752050, India}
\ead{jk14@iitbbs.ac.in, ag34@iitbbs.ac.in, malay@iitbbs.ac.in}
\vspace{10pt}

\begin{indented}
\item[19th January 2022]
\end{indented}

\begin{abstract}
In this brief report, following the recent developments on formulating a quantum analogue of the classical energy equipartition theorem for open systems where the heat bath comprises of independent oscillators, i.e. bosonic degrees of freedom, we present an analogous result for fermionic systems. The most general case where the system is connected to multiple reservoirs is considered and the mean energy in the steady state is expressed as an integral over the reservoir frequencies. Physically this would correspond to summing over the contributions of the bath degrees of freedom to the mean energy of the system over a suitable distribution function \(\rho(\omega)\) dependent on the system parameters. This result holds for nonequilibrium steady states, even in the nonlinear regime far from equilibrium. We also analyze the zero temperature behaviour and low temperature corrections to the mean energy of the system.
\end{abstract}

%
%
%
%
%

\section{Introduction}
In recent years, it has been demonstrated that the laws of quantum mechanics are consistent with those of thermodynamics. This is rather surprising because quantum systems are nowhere close to the thermodynamic limit and may even consist of a single particle. This remarkable aspect of quantum systems has fuelled a considerable amount of recent developments. For example, several equilibrium properties of generic quantum systems have been investigated using both the Langevin equation and the Gibbs ensemble methods \cite{QLE,qt1,qt2,qt3,qt4,qt5,qt6}. Quite notable is the study of nonequilibrium steady states \cite{ness1,ness2} for such systems. It turns out that for certain systems in the nanoscale or mesoscopic regimes, it is possible to express the reduced density matrix in the form of a generalized exponential, the so called McLennan–Zubarev form (see also \cite{MZ1,MZ2}). This has allowed a formulation of nonequilibrium steady states in a manner analogous to equilibrium thermodynamics with the introduction of generalized Massieu-Planck potentials \cite{ness1}.\\

Recently there has been a considerable amount of interest in studying the quantum counterpart of the energy equipartition theorem for open quantum systems \cite{jarzy1,jarzy2,jarzy3,jarzy4,jarzy5,kaur}. In particular, it has been demonstrated that in the steady state, the average energy \(E\) of an open quantum system receives contributions from the bath degrees of freedom and can typically be cast in the form, \(E =\) \textit{sum of contributions from the degrees of freedom of the bath}. The manner in which the bath degrees of freedom contribute to the mean energy of the system of interest depends on several control factors such as the dissipation mechanism, memory time and externally applied fields \& potentials \cite{jarzy2,kaur}. However, in all the previous studies on this subject so far, the bath with which the system is kept in contact is taken to be composed of independent harmonic oscillators (see for example \cite{jarzy4}). This is because the bath degrees of freedom are taken to be harmonic oscillators with an implicit bosonic character. In such a setting the mean energy of the system is expressible in the following exact form,
\begin{equation}\label{123}
  E = \int_{0}^{\infty} \mathcal{E}(\omega,T) \mathcal{P}(\omega) d\omega
\end{equation} where \(\mathcal{E}(\omega,T)\) is the mean energy of an individual bath oscillator with frequency \(\omega\) and kept at temperature \(T\) while \(\mathcal{P}(\omega)\) is a suitable function of the bath frequencies. In fact, it can be shown that \(\mathcal{P}(\omega)\) is a probability distribution function \cite{jarzy4,kaur} meaning that the bath oscillators of various frequencies contribute to the mean energy of the system according to probabilities dictated by \(\mathcal{P}(\omega)\).\\

In the present paper, we generalize the quantum counterpart of energy equipartition theorem for generic fermionic systems. For the sake of generality, we take the system to be connected to multiple noninteracting reservoirs. Such a coupling is in general strong and in the steady state, the system might as well be far from equilibrium. Recently, it has been observed that thermodynamic
laws are very much compatible with quantum properties of open nanoscale
systems comprised of multiple reservoirs with
different chemical potentials and temperatures (see figure-(1)) \cite{ness1,ness2}. Although such systems involve only few particles, the emergence of thermodynamics laws is also related to quantum
nature, and the emergence of such thermodynamics results may require averaging over reservoir degrees of freedom \cite{ness1}. Henceforth, we consider such nanoscale systems which open the doorway of a rare and
novel opportunity to study the steady-state quantum thermodynamics, without much dependence on usual statistical ensemble hypothesis \cite{ness1}. Furthermore, our final results are fairly general and robust as long as one is in a nonequilibrium steady state. The mean energy of the system is expressed as a two fold average just as in the case of the dissipative oscillator (see \cite{jarzy3}) where for the latter, \(\mathcal{E}(\omega,T)\) in eqn (\ref{123}) is the mean energy of an individual thermostat oscillator of frequency \(\omega\) and at temperature \(T\) obtained by averaging over the Gibbs canonical state of the bath. The second averaging occurs in eqn (\ref{123}) as an average over the bath frequencies \(\omega\).\\

The manuscript is organized as follows. In the next section, we present the model set-up and obtain the rather familiar expression for the mean energy of the system as an integral over the reservoir spectrum. In section 3, we propose and analyze the quantum counterpart of energy equipartition theorem for our general nanoscale fermionic system. The zero temperature and low temperature corrections to the mean energy are also discussed. Although our results in section 3 are based on wide-band approximation, but we generalize it for more general bath spectrum in section 4. We conclude our paper in the final section.

\section{Model \& preliminaries}
\begin{figure}
\begin{center}
\includegraphics[scale=0.52]{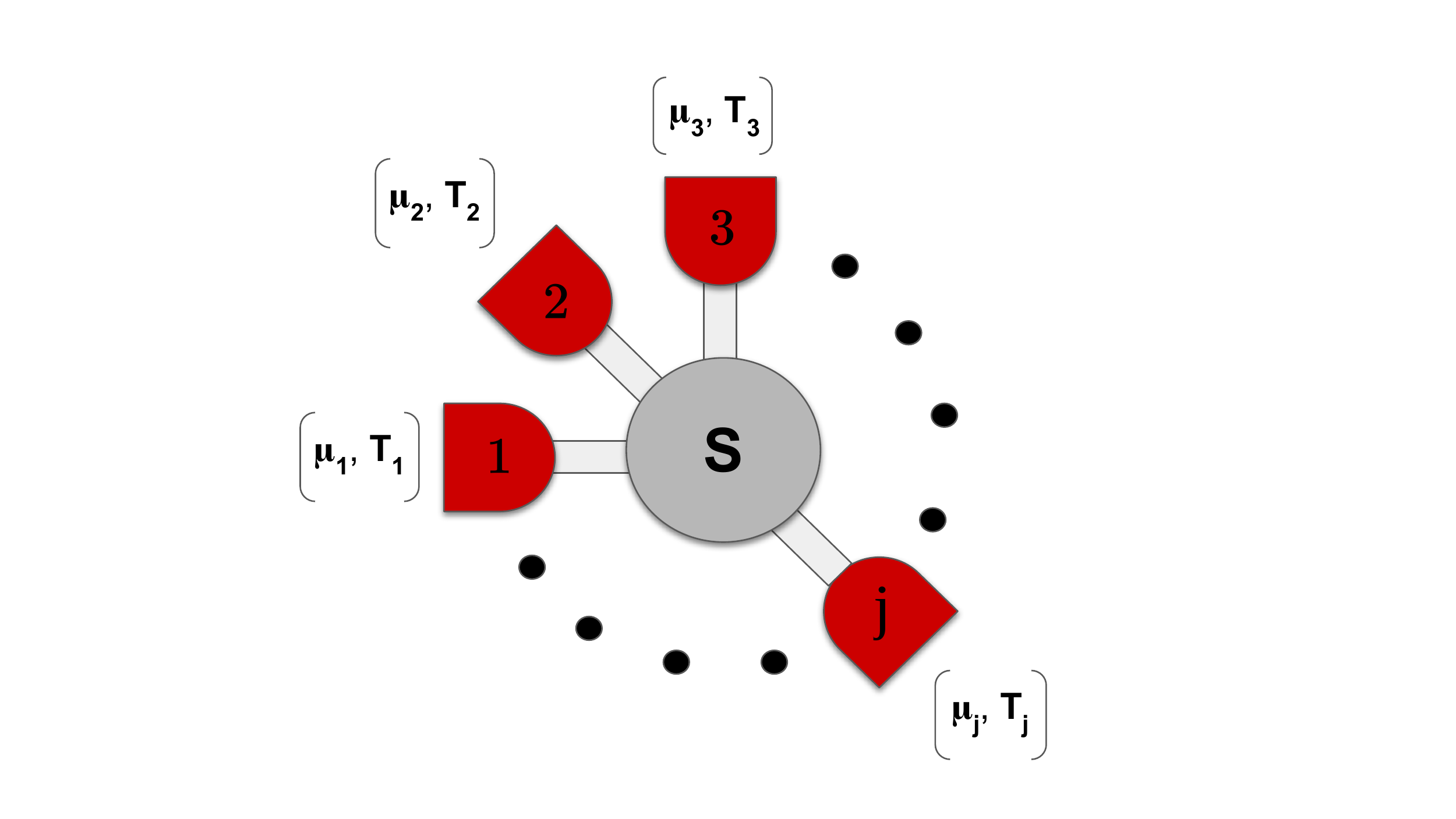}
\caption{Our general model system consists of a quantum dot connected with multiple fermionic reservoirs with different temperatures ($T_j$) and chemical potentials ($\mu_j$) for $j=1,2,3,.....N$.}
\label{model}
\end{center}
\end{figure}
The system of interest here is a quantum dot in contact with several fermionic reservoirs maintained at different temperatures ($T_j$) and chemical potentials ($\mu_j$). The total Hamiltonian consists of three parts,
\begin{equation}
H = H_S + H_B + H_{SB}
\end{equation}
where $H_S$ is the Hamiltonian for the system of a single dot (the “subsystem”), $H_B$
includes several metallic leads at different temperatures and chemical potentials, and $H_{SB}$ incorporates subsystem-bath hybridization terms. Specifically, we can mention that the subsystem is described by,
\begin{equation}
 H_{S} = \omega_s a^\dagger a
\end{equation}
 where \(\omega_s\) is the subsystem energy (since we consider $\hbar=1$), $a^\dagger$ and $a$ are the subsystem creation and annihilation operators. Further, the metallic leads are composed of noninteracting electrons and they are expressed as,
 \begin{equation}
 H_B= \sum_{j=1}^{N}H_{B,j}=\sum_{j=1}^{N}\sum_{k=1}^{\infty}\omega_{j,k}b_{j,k}^\dagger b_{j,k}
 \end{equation}
 where $b_{j,k}^\dagger$ and $b_{j,k}$ are the creation and annihilation operators for the $k$th state with energy $\omega_{j,k}$ of the $j$th metallic lead. Hamiltonian for the $j$th lead is denoted by $(H_{B,j}$ . The Hamiltonian summarizing the hybridization between the subsystem and the metallic leads is given by,
\begin{equation}
  H_{SB} = \sum_{j} \sum_{k} [\xi_{j, k} a^\dagger b_{j,k} + {\rm h. c.}]
\end{equation}
where $\xi_{j, k}$ is the coupling strength between the $k$th state of the $j$th metallic lead and the subsystem. One can notice that the coupling is bilinear in nature. Such a Hamiltonian can be derived from a fully coupled (FC) system-bath model (see also \cite{FC}) by considering the rotating wave approximation (RWA) wherein the rapidly oscillating terms are omitted \cite{QLE}. For a fully coupled (FC) Hamiltonian, the interaction part contains four terms, namely $a^{\dagger}b_{j,k}$, $b_{j,k}^{\dagger}a$, $ab_{j,k}$ and $a^{\dagger}b_{j,k}^{\dagger}$. The first two terms correspond to real processes conserving the unperturbed energy, while the other two are known as the counter-rotating terms. The last two terms describe events not corresponding to real absorption and emission processes because of which they are called virtual processes. In the second order in perturbation theory, both the two real and virtual processes combine to give rise to real processes. Although these two different coupling models lead to different short time behaviours, in the asymptotic long time Markovian regime the two actually coincide. In the present paper our study is related with equipartition theorem which is obtained in the asymptotic long time regime. Henceforth, we believe that our results are not affected by changing our coupling scheme from RWA to FC.\\

We are interested in fermionic degrees of freedom, thus all the creation and annihilation operators obey anti-commutation relations given by,
\begin{equation}
\{a,a^\dagger\} = 1, \hspace{5mm} \{ b_{j,k}, b_{j',k'}^\dagger\} = \delta_{jj'} \delta_{kk'}
\end{equation} with all others vanishing. It needs to be mentioned here that the presence of the reservoir makes the nanoscale system dissipative, and they induce a finite resonant width $\gamma_j = \sum_k\pi|\xi_{j,k}|^2\rho_j$
due to the reservoir $j$ (with its density of states $\rho_j$).\\

Since our model set-up is noninteracting, its steady-state characteristics
can be expressed exactly in terms of the nonequilibrium
Green’s function (NEGF) approach \cite{mb1,mb2}. The present derivation follows an equation-of-motion approach \cite{mb1,mb2}. Here, one can obtain an effective quantum
Langevin equation for the subsystem by solving the
Heisenberg equations of motion (EOM) for the bath variables and subsequently plugging them back
into the EOM for the subsystem (dot) variables. The resulting quantum Langevin equation reads the following,
\begin{eqnarray}
   \frac{da(t)}{dt} &+& i \int_{t_0}^{t} dt' \sum_{j} \sum_k \xi_{j, k} g_{j,k}^+(t - t') \xi^*_{j,k} a(t') +i \omega_s a(t)  = i \sum_j \eta_j(t) \label{eoma}
\end{eqnarray}
where \(g_{j,k}^+(t)\) refers to the retarded Green's function of the  \(j\)th isolated reservoir, being given as,
\begin{equation}
  g_{j,k}^+(t)= - i e^{-i \omega_{j,k} t} \theta(t)
\end{equation} while \(\eta_j(t)\) refers to the quantum noise due to the \(j\)th bath which takes the following explicit form,
\begin{equation}
  \eta_j(t) = -i \sum_k \xi_{j, k} g_{j,k}^+(t-t_0) b_{j,k}(t_0).
\end{equation}
Now eqn (\ref{eoma}) can be solved in the Fourier domain as,
\begin{equation}\label{solution}
  \tilde{a}(\omega) = G^+(\omega)  \sum_{j}  \tilde{\eta}_{j}(\omega)
\end{equation} where \(G^+(\omega)\) is the Fourier transformed retarded Green's function. For the present case, it reads,
\begin{equation}
  G^+(\omega) = \frac{1}{(\omega - \omega_s)- \sum_{j} \sum_{k} \xi_{j, k} \tilde{g}^+_{j,k}(\omega) \xi^*_{j,k}}.
\end{equation}
Here, \(\sum_{k} \xi_{j, k} \tilde{g}^+_{j,k}(\omega) \xi^*_{j,k}\) refers to the self energy of the \(j\)th bath. The real part of the self-energy is a principal value integral, which vanishes when the bath's density of states is energy independent and the bandwidth is large. This is well known as the wide-band approximation \cite{landauer,imry,wingreen}.
To this end, we define the hybridization strengths between the system and the baths as,
\begin{equation}\label{Gamma}
  \Gamma_j = 2\pi \sum_{k} \xi_{j, k} \delta(\omega - \omega_{j,k}) \xi^*_{j,k}.
\end{equation}
With these definitions, the reduced density matrix (in the frequency domain) of the subsystem in the steady state reads \cite{mb1,saito}, 
\begin{equation}\label{reducedmatrixj}
\mathfrak{P}^S(\omega) d\omega = \langle \tilde{a}^{\dagger}(\omega)\tilde{a}(\omega) \rangle d \omega =  \frac{1}{2\pi} \bigg[\sum_j G^{+}(\omega)\Gamma_j (\omega) G^{-}(\omega)f(\omega,\mu_j,T_j)\bigg] d\omega
\end{equation} where  \(G^{-} = (G^{+})^\dagger\) and we have used eqn (\ref{solution}) along with the correlation function of the noise operators $\langle \tilde{\eta}_j^{\dagger}(\omega)\tilde{\eta}_k(\omega) \rangle =\frac{\Gamma_{k}}{2\pi}\delta_{j,k}f(\omega,\mu_k,T_k)$. In the above equation, \(f(\omega,\mu_j,T_j)\) is the Fermi distribution function for the \(j\)th bath. Thus, the mean energy of the subsystem in the frequency interval \(\omega\) to \(\omega + d\omega\) can be expressed in terms of the reduced density matrix [eqn (\ref{reducedmatrixj})] as $\epsilon(\omega) d\omega =\omega \mathfrak{P}^S(\omega)d\omega$. This implies that the mean energy of the subsystem (the quantum dot) when it interacts with a number of non-interacting baths has the following form in the steady state, 
\begin{equation}\label{avgenergy}
E =\sum_{j}E_j = \frac{1}{2\pi} \int_{-\infty}^{\infty}  \bigg[\sum_j G^{+}(\omega) \Gamma_j (\omega) G^{-}(\omega) f(\omega,\mu_j,T_j)\bigg] \omega d\omega
\end{equation} where, 
\begin{equation}
E_j = \frac{1}{2\pi} \int_{-\infty}^{\infty}  \bigg[G^{+} (\omega) \Gamma_j (\omega) G^{-}(\omega) f(\omega,\mu_j,T_j)\bigg] \omega d\omega
\end{equation} is the contribution due to the \(j\)th bath. In the wide-band approximation, we take \(\xi_{j,k}\) to be a real constant independent of other parameters. The hybridization strengths are therefore \(\Gamma_j := \gamma_j\) which are real constants and consequently eqn (\ref{avgenergy}) reads (also see \cite{ness1}),
\begin{equation}\label{avgenergy1}
  E = \int_{-\infty}^{\infty} \bigg[\sum_j \frac{\gamma_j}{\gamma} f(\omega,\mu_j,T_j)\bigg] \omega \rho(\omega) d\omega
\end{equation}
where \(\gamma = \sum_j \gamma_j\) and \(\rho(\omega)\) is the spectral function of the system,
\begin{equation}\label{rhoonebath}
  \rho(\omega) = \frac{\gamma}{\pi} \frac{1}{(\omega - \omega_s)^2 + \gamma^2}.
\end{equation}
It should be remarked that although in general there is a non-trivial energy and particle transport taking place through the system (the quantum dot), the expression for the mean energy [eqn (\ref{avgenergy1})] holds good. The above equation is straightforwardly generalized to cases where there the subsystem is connected to multiple external baths. As such, the analogous expression for \(\rho(\omega)\) is,
\begin{equation}\label{rhomanybath}
  \rho(\omega) = \frac{\tilde{\gamma}}{\pi} \frac{1}{(\omega - \omega_s)^2 + \tilde{\gamma}^2}
\end{equation} where \(\tilde{\gamma} = \sum_j \gamma_j\) and \(\gamma_j\) is the hybridization strength for the \(j\)th bath.

\section{Energy partition in fermionic systems}
We can now propose a quantum counterpart for the energy equipartition theorem for fermionic systems along similar lines as refs \cite{jarzy1,jarzy2,jarzy3,jarzy4,kaur}. Since \(f(\omega,\mu_j,T_j)\) is the distribution function for the \(j\)th bath, i.e. \(f(\omega,\mu_j,T_j) := \langle b^\dagger_{j,k} b_{j,k} \rangle\), one can identify \(\omega f(\omega,\mu_j,T_j)\) as the mean energy of the bath degrees of freedom of the \(j\)th bath lying in the range \(\omega\) to \(\omega + d\omega\). Let us denote this as \(\mathcal{E}(\omega,\mu_j,T_j) = \omega f(\omega,\mu_j,T_j)\). Subsequently, eqn (\ref{avgenergy1}) can be re-written as,
\begin{equation}\label{qceeth}
  E = \int_{-\infty}^{\infty} \bigg[\sum_j \frac{\gamma_j}{\gamma} \mathcal{E}(\omega,\mu_j,T_j)\bigg] \rho(\omega) d\omega.
\end{equation}
In other words, the mean energy of the quantum system can be expressed as a sum over contributions received from the heat bath degrees of freedom over the entire frequency spectrum. Such contributions are controlled by a suitable distribution function. For simplicity, if one considers a single heat bath, eqn (\ref{qceeth}) reads,
\begin{equation}\label{equipartition}
  E = \int_{-\infty}^{\infty} \mathcal{E}(\omega,\mu,T) \rho(\omega) d\omega
\end{equation} which appears similar to the quantum counterpart of energy equipartition theorem as first proposed in \cite{jarzy1}. Let us remind ourselves that the net energy is expressed as a two fold average: the first one is over the Gibbs state of the fermionic quantum heat bath yielding \(\mathcal{E}(\omega,\mu,T)\) as the mean energy of the degrees of freedom lying in the range \(\omega\) to \(\omega + d\omega\) whereas, the second averaging takes place over the distribution function \(\rho(\omega)\) which dictates the extent of contributions received from a particular frequency from the bath spectrum.\\

\begin{figure}
\begin{center}
\includegraphics[scale=0.70]{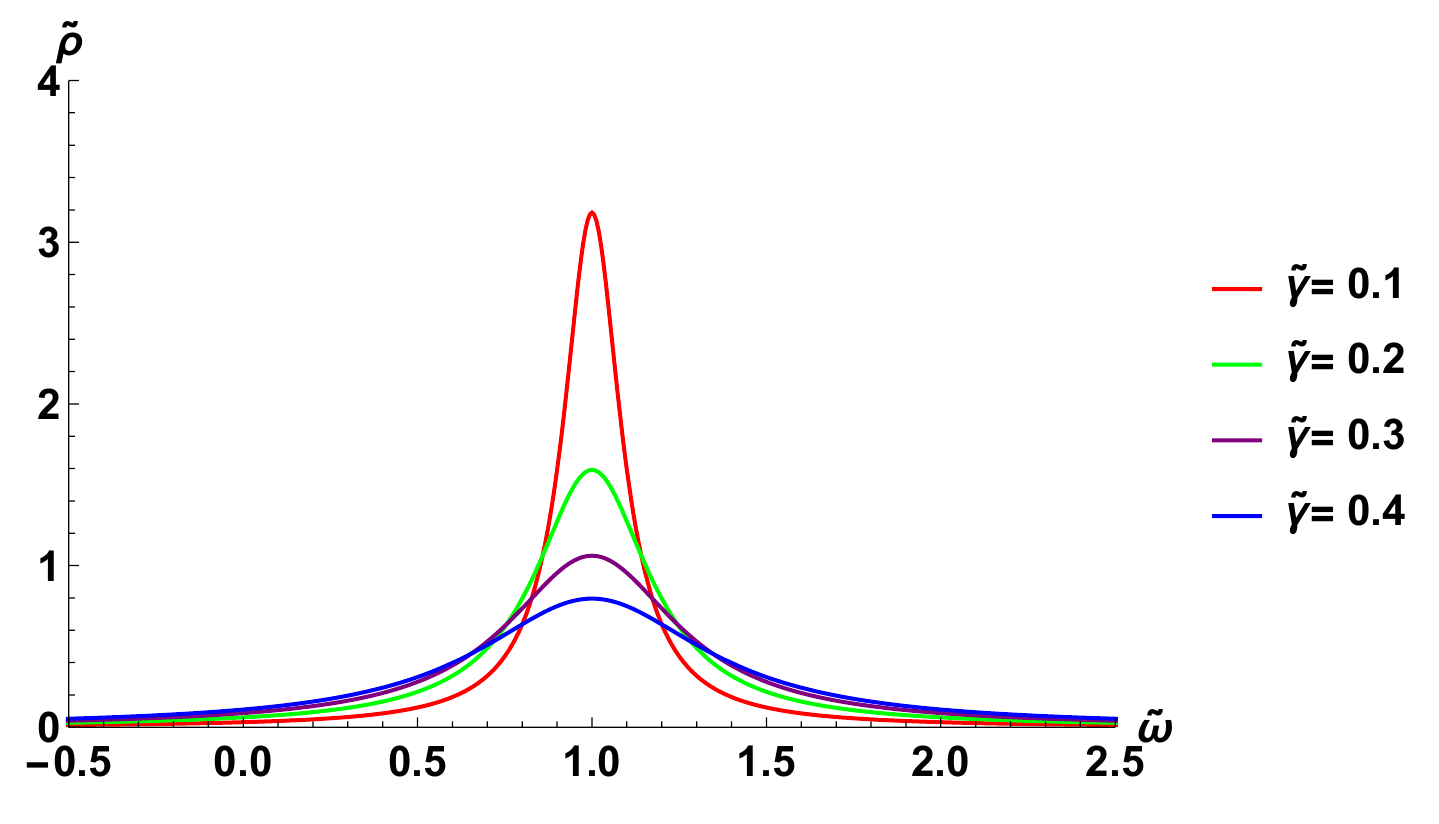}
\caption{Variation of \(\tilde{\rho}(\tilde{\omega}) = \omega_s \rho(\omega/\omega_s)\) as a function of the dimensionless bath frequencies \(\tilde{\omega} = \omega/\omega_s\) for different values of \(\tilde{\gamma} = \gamma/\omega_s\).}
\label{tilde rho vs tilde omega}
\end{center}
\end{figure}

Following the same arguments as that of ref \cite{kaur}, it is straightforward to check that \(\rho(\omega)\) is positive semi-definite, i.e. \(\rho(\omega) \geq 0\) for all values of \(\omega\) and that,
\begin{equation}
  \int_{-\infty}^{\infty} \rho(\omega) d\omega =1
\end{equation}  it is also normalized. Thus, \(\rho(\omega)\) can be regarded to be a probability distribution function analogous to the one defined in \cite{jarzy1,jarzy2,jarzy3,jarzy4,kaur} where the degrees of freedom were bosonic. The distribution function \(\rho(\omega)\) for a single bath has been plotted in figure-(\ref{tilde rho vs tilde omega}). The influence of the system-bath hybridization strength \(\gamma\) and the system energy parameter \(\omega_s\) is clearly demonstrated. These parameters are analogous to the damping strength and harmonic trap frequency respectively in the case of a dissipative oscillator (for example, see \cite{jarzy3}) where the degrees of freedom are bosonic by definition. From figure-(\ref{tilde rho vs tilde omega}), it appears as if in the nonequilibrium steady state, the system receives most contributions to the mean energy from reservoir frequencies lying around \(\omega_s\). Further, the smaller the hybridization strength is, the lesser are the contributions received from frequencies away from \(\omega_s\).\\

For the case of multiple baths being connected to the subsystem, one can write the mean energy as a simple superposition of contributions received from different baths weighted by their hybridization strengths as,
\begin{equation}
  E = \sum_{j} \frac{\gamma_j}{\gamma} \int_{-\infty}^{\infty} \rho(\omega) \mathcal{E}(\omega,\mu_j,T_j) d\omega.
\end{equation}
Perhaps the most elegant aspect of the above result is that the mean energy turns out to be a linear sum of contributions coming from each bath. This kind of structure results in the thermodynamic Massieu-Planck potential of the overall subsystem to be written as a sum of single reservoir Massieu-Planck potentials \cite{ness1}. The mean energy can be computed by evaluating the integral over the bath frequencies. However, in the wide-band approximation as invoked above, the integral diverges. This is analogous to the case of strict Ohmic dissipation where the friction kernel in the frequency domain is a real constant, independent of the bath frequencies. However, the integral above can be performed via a suitable regularization such as introducing a finite bandwidth for the bath(s). The reader is referred to the recent work \cite{ness1} (see appendix C therein) for the details. We now analyze the low temperature features of the mean energy.

\subsection{Zero temperature case}
Consider the case of a single heat bath. If we set \(T = 0\) and let \(\mu_0\) be the chemical potential at zero temperature (the Fermi energy), the mean energy of the subsystem reads,
\begin{equation}\label{T0}
  E_{T = 0} = \int_{-\infty}^{\mu_0} \omega \rho(\omega) d\omega.
\end{equation}
This integral diverges due to the lower limit and therefore we introduce a lower cutoff frequency to regularize the integral, i.e. we write,
\begin{equation}\label{T0}
  E_{T = 0} =  \frac{\gamma}{\pi} \int_{-\omega_{\rm cut}}^{\mu_0} \frac{\omega d\omega}{(\omega - \omega_s)^2 + \gamma^2}
\end{equation}
which can be re-expressed as,
\begin{eqnarray}
  E_{T = 0} = \frac{\gamma \omega_s}{\pi} \int_{-\omega_{\rm cut}}^{\mu_0} \frac{d\omega}{(\omega - \omega_s)^2 + \gamma^2}
   + \frac{\gamma}{2 \pi} \int_{-\omega_{\rm cut}}^{\mu_0} \frac{d[(\omega - \omega_s)^2 + \gamma^2]}{(\omega - \omega_s)^2 + \gamma^2}.
\end{eqnarray}
 This yields the final answer for the zero temperature mean energy,
 \begin{eqnarray}
   E_{T = 0} &=& \frac{\omega_s}{\pi} \tan ^{-1} \bigg[ \frac{\gamma(\mu_0 + \omega_{\rm cut})}{\gamma^2 - (\mu_0 - \omega_s)(\omega_{\rm cut} + \omega_s)} \bigg] \nonumber \\
    &+& \frac{\gamma}{2 \pi} \ln \bigg[ \frac{(\mu_0 - \omega_s)^2 + \gamma^2}{(\omega_{\rm cut} + \omega_s)^2 + \gamma^2} \bigg]. \label{ET0}
 \end{eqnarray}
In the weak coupling regime, one has \(\gamma \rightarrow 0 \), i.e. the quantum dot and the metallic lead (the bath) are weakly coupled. In this limit, using the identity $\lim_{y \rightarrow 0}\frac{y}{x^2+y^2}=\delta(x)$, Eq. (\ref{T0}) reduces to,
\begin{equation}
E_{T = 0} =   \int_{-\omega_{\rm cut}}^{\mu_0} \omega \delta (\omega - \omega_s) d\omega.
\end{equation} Since \(\omega_s \in [- \omega_{\rm cut}, \mu_0]\), this gives \(E_{T = 0} = \omega_s\) in the weak coupling (\(\gamma \rightarrow 0\)) limit. This is expected because if one removes the effect of an external bath, the mean energy of the dot should coincide with the system characteristic energy \(\omega_s\).\\

One may also discuss the strong coupling regime wherein, one has \(\gamma \rightarrow \infty\). Thus, one may re-write the integrand of eqn (\ref{T0}) as follows,
\begin{equation}
\frac{\gamma}{\pi} \frac{\omega}{(\omega - \omega_s)^2 + \gamma^2} = \frac{\omega}{\gamma \pi} \Bigg[ 1 + \bigg(\frac{\omega - \omega_s}{\gamma}\bigg)^2 \Bigg]^{-1}  \approx \frac{\omega}{\gamma \pi} \Bigg[ 1 - \bigg(\frac{\omega - \omega_s}{\gamma}\bigg)^2 \Bigg].
\end{equation}
 Plugging this into the integral, one can obtain the form of \(E_{T = 0}\) as a function of the hybridization strength as,
 \begin{equation}
E_{T = 0} \approx \frac{A}{\gamma} + \frac{B}{\gamma^3}
\end{equation} where the constants \(A=\frac{(\mu_0^2-\omega_{\rm cut}^2)}{2\pi}\) and \(B=\frac{2\omega_s(\mu_0^{3}+\omega_{\rm cut}^{3})}{3\pi}+\frac{\omega_s^2(\omega_{\rm cut}^2-\mu_0^2)}{2\pi}+\frac{(\omega_{\rm cut}^4-\mu_0^4)}{4\pi}\) depend on the parameters \(\mu_0\), \(\omega_s\) and \(\omega_{\rm cut}\). Thus it is clear that as \(\gamma \rightarrow \infty\), the zero temperature mean energy of the dot goes to zero as  \(E_{T = 0} \sim 1/\gamma\). This is in sharp contrast to the case of the free Brownian particle considered in \cite{jarzy5} wherein for the case of Drude dissipation, the zero temperature mean energy of the particle goes to infinity as the coupling strength tends to infinity.
\subsection{Low temperature corrections}
\begin{figure}
\begin{center}
\includegraphics[scale=0.60]{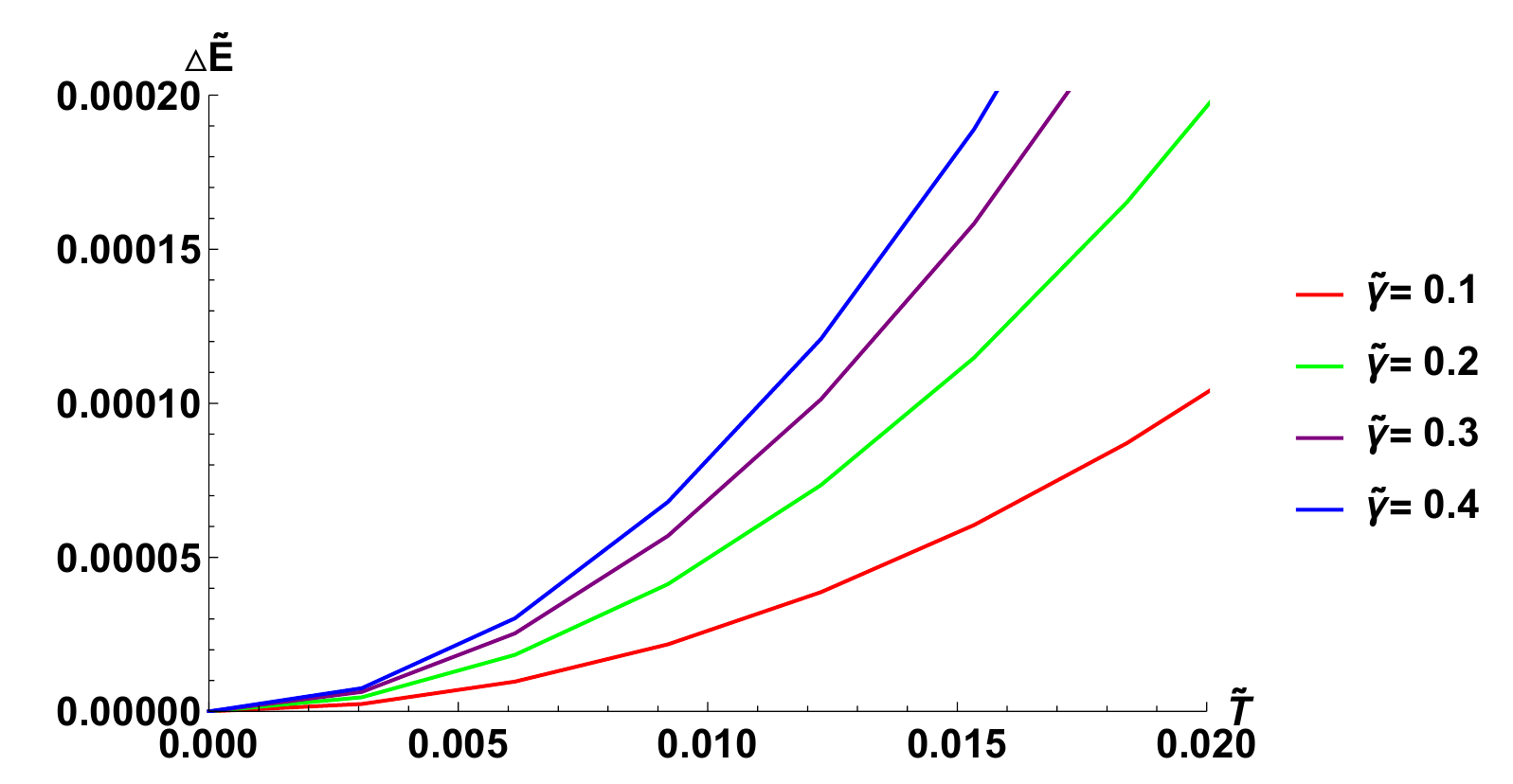}
\caption{Variation of \(\Delta \tilde{E} = (E_{T \rightarrow 0} - E_{T = 0})/\omega_s\) as a function of \(\tilde{T} = k_B T/\omega_s\) for different values of \(\tilde{\gamma} = \gamma/\omega_s\).}\label{tilde E vs Tilde T}
\includegraphics[scale=0.60]{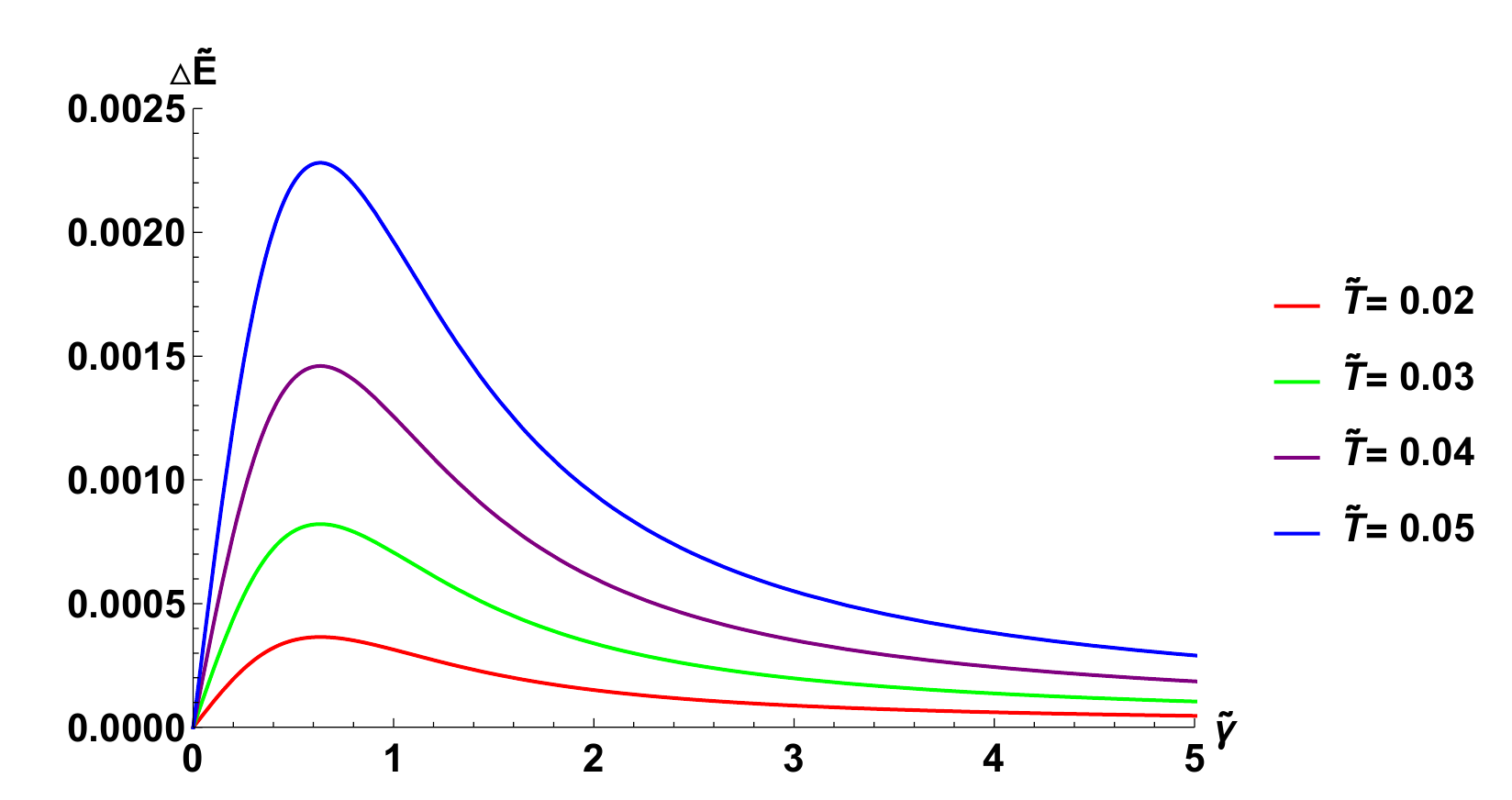}
\caption{Variation of \(\Delta \tilde{E} = (E_{T \rightarrow 0} - E_{T = 0})/\omega_s\) as a function of \(\tilde{\gamma} = \gamma/\omega_s\) for different values of \(\tilde{T} = k_B T/\omega_s\).}\label{TILDE E vs Tilde gamma}
\end{center}
\end{figure}

Now that we have discussed the zero temperature case, we analyze the low temperature corrections to the mean energy of the subsystem. Noting that \(\omega \rho(\omega)\) vanishes as \(\omega \rightarrow \pm \infty\), we can employ the Sommerfeld expansion to yield the low temperature corrections to mean energy,
 \begin{eqnarray}
   E_{T \rightarrow 0} &=& \lim_{T \rightarrow 0} \int_{-\infty}^{\infty} \omega \rho(\omega) f(\omega,\mu,T) d\omega \nonumber \\
    &\approx&  \int_{-\infty}^{\mu_0} \omega \rho(\omega) d\omega + \frac{\pi^2}{6} (k_B T)^2 \big[\rho(\mu_0) + \mu_0 \rho'(\mu_0)\big] \nonumber \\
 \end{eqnarray}
 Note that the first term is simply the zero temperature contribution \(E_{T = 0}\). Thus, the low temperature corrections are,
 \begin{equation}\label{corrections}
   E_{T \rightarrow 0} - E_{T = 0} = \frac{\pi \gamma}{6} (k_B T)^2 \Bigg[\frac{(3\mu_0 - \omega_s)(\mu_0 - \omega_s) + \gamma^2}{\big[ (\mu_0 - \omega_s)^2 + \gamma^2\big]^2}\Bigg].
 \end{equation} The corrections are plotted in figures-(\ref{tilde E vs Tilde T}) and (\ref{TILDE E vs Tilde gamma}). The dependence on temperature is quadratic as can be clearly seen from eqn (\ref{corrections}). The manner in which these corrections depend on the subsystem-bath hybridization strength is however, somewhat nontrivial as can be seen from figure-(\ref{TILDE E vs Tilde gamma}). These calculations can be straightforwardly generalized for cases where the subsystem is in contact with multiple reservoirs.

\section{Beyond the wide-band approximation}
In the previous section, we have established the quantum counterpart of the energy equipartition theorem for a generic fermionic system where the subsystem is in a nonequilibrium steady state and can be in contact with multiple external reservoirs. However, for the sake of simplicity, we had invoked the wide-band approximation wherein the subsystem-bath coupling is taken to be independent of the reservoir levels so that the hybridization strengths are real constants. In the more general setting, where this approximation may not necessarily hold eqn (\ref{avgenergy}) is still valid. It is then natural to wonder as to whether one still has a quantum counterpart of the equipartition theorem. For simplicity, let us consider the case of a single reservoir connected to the subsystem. A generalization to multiple reservoirs can be performed easily. If we identify the function \(2 \pi \rho(\omega) = G^+ \Gamma G^-\), then eqn (\ref{avgenergy}) is identical in structure to eqn (\ref{equipartition}) which is the quantum counterpart of the equipartition theorem. It then remains to show whether the function \(\rho(\omega)\) satisfies the basic properties of a probability distribution function, i.e. it is positive semi-definite and is normalized.\\

Positivity of \(\rho(\omega)\) can be easily demonstrated. Noting that the energy [eqn (\ref{avgenergy})] is real, i.e. is equal to its complex conjugate, one must have \(\Gamma(\omega)\) to be a real valued function. Furthermore, referring to eqn (\ref{Gamma}) where \(\Gamma\) is defined in terms of the microscopic coupling parameters, one finds that \(\Gamma(\omega) \geq 0\) \(\forall\) \(\omega \in (-\infty,\infty)\). Note that this does not impose any restriction on the coupling parameters \(\{\xi_{j,k}\}\): they may still be complex and are in general frequency dependent. Therefore, \(\rho(\omega)\), which has the following expression (using the fact the self energy \(\sum_{k} \xi_{k} \tilde{g}^+_{k}(\omega) \xi^*_{k} = - i \Gamma\)),
\begin{equation}
  \rho(\omega) = \frac{1}{ \pi}\frac{\Gamma(\omega)}{(\omega - \omega_s)^2 + \Gamma(\omega)^2}
\end{equation} is positive semi-definite or \(\rho(\omega) \geq 0\) \(\forall\) \(\omega \in (-\infty,\infty)\).\\

Next, one has to show that \(\rho(\omega)\) is normalized, i.e.
\begin{equation}
  \int_{-\infty}^{\infty} \frac{\Gamma(\omega)}{(\omega - \omega_s)^2 + \Gamma(\omega)^2} = \pi.
\end{equation} Let us begin by noting that although \(\Gamma(\omega)\) may have poles, they do not affect the integral because the function appears in the denominator with a larger power than it appears in the numerator. However, its zeroes play a non-trivial role in the integral. Making use of the identity,
\begin{equation}
  \lim_{\epsilon \rightarrow 0} \frac{\epsilon}{(x-a)^2 + \epsilon^2} = \pi \delta(x-a)
\end{equation} one easily obtains,
\begin{equation}
  \int_{-\infty}^{\infty} \rho(\omega) d\omega = \frac{1}{\pi} \int_{-\infty}^{\infty} \pi \delta(\omega - \omega_s) = 1
\end{equation} thus verifying that \(\rho(\omega)\) is normalized. Therefore, in other words, even in a more general setting where the subsystem-bath hybridization strength is frequency dependent, the quantum counterpart of energy equipartition theorem holds good and the mean energy of the subsystem can be expressed as a two fold average.

\section{Conclusion}
In this paper, we have formulated a quantum analogue of the classical equipartition theorem for a quantum dot connected with multiple metallic leads at different temperatures and chemical potentials. Henceforth, our subsystem is connected to multiple reservoirs and we have developed the quantum analogue of energy equipartition theorem for a general open fermionic system at a nonequilibrium steady state. The mean energy is expressible in the form of a two fold average. The first averaging takes place over the thermal state of the reservoir while the second is taken over the entire reservoir spectrum wherein contributions to the subsystem's mean energy from the bath degrees of freedom of different frequencies are incorporated. It should perhaps be strongly emphasized that this result does not depend on whether the system is close to thermodynamic equilibrium or not as long as the system is described by a nonequilibrium steady state. In other words, even if the system is in the strong nonlinear regime with nonlinear energy and particle currents passing through it, the proposed quantum counterpart of energy equipartition theorem holds good.\\

It should be noted that although the equipartition theorem and its quantum counterpart is discussed in the literature typically in the context of the kinetic energy (and sometimes the potential energy), in the present case, the subsystem energy does not have a clear identification to kinetic and potential energies. Naively, it may be interpreted as the mean energy in the steady state contained in the subsystem (the quantum dot) given that it is interacting with multiple non-interacting baths. If on the other hand, one thinks of such a system as a lattice, the mean energy can be interpreted as the mean site energy.\\

One may observe that although we have studied the fermionic case, the results are quite analogous to those for bosonic systems (with the creation and annihilation operators obeying commutation relations) as obtained in ref. \cite{kaur}. Our approach therefore leads to a rather general statement about the quantum thermodynamics of nanoscale systems. Furthermore, one can observe that the quantum probability distributions associated to the quantum equipartition theorem can distinguish the properties of the quantum environment and its coupling to a given quantum system, and hence, they may be experimentally quantified from the measurement of the linear response of the system to an applied  perturbation, for instance, electrical or magnetic. In particular, our model system may open a pathway to investigate the influence of various dissipation mechanisms, external magnetic field, the confining potential strength, and the memory time on the average energy of an open fermionic system.

\ack
J.K. acknowledges the financial support received from IIT Bhubaneswar in the form of an Institute Research Fellowship. The work of A.G. is supported by the M.H.R.D., Government of India in the form of a PMRF. M.B. gratefully acknowledges financial support from Department of Science and Technology (DST), India under the Core grant (Project No. CRG/2020//001768).

\end{document}